\documentclass[aps,twocolumn,preprintnumbers,showpacs,superscriptaddress,%groupedaddress,
  nofootinbib,floatfix]{revtex4-1} 

\pdfoutput=1

\usepackage[T1]{fontenc} 
\usepackage[utf8]{inputenc}
\usepackage{booktabs}
\usepackage{amsmath}
\usepackage{amssymb}
\usepackage{graphicx}
\usepackage{multirow}
\usepackage{xcolor}

\begin{document}

\title{Astrophysics-independent determination of dark matter parameters
  from two direct detection signals}

\author{Juan Herrero-Garc\'{\i}a}
\email{jherrero@sissa.it}
\affiliation{SISSA/INFN, Via Bonomea 265, I-34136 Trieste, Italy}

\author{Yannick M\"uller}
\email{ym-steinweiler@gmx.de}
\affiliation{Institute of Nuclear Physics, Karlsruhe Institute of Technology (KIT)\\
 Hermann-von-Helmholtz-Platz 1, 76344 Eggenstein-Leopoldshafen, Germany}

\author{Thomas Schwetz}
\email{schwetz@kit.edu}
\affiliation{Institute of Nuclear Physics, Karlsruhe Institute of Technology (KIT)\\
 Hermann-von-Helmholtz-Platz 1, 76344 Eggenstein-Leopoldshafen, Germany}%

\preprint{SISSA  22/2019/FISI}

\begin{abstract}
Next-generation dark matter direct detection experiments will explore several orders of magnitude in the dark matter--nucleus scattering cross section below current upper limits. In case a signal is discovered the immediate task will be to determine the dark matter mass and to study the underlying interactions. We develop a framework to determine the dark matter mass from signals in two experiments with different targets, independent of astrophysics. Our method relies on a distribution-free, nonparametric two-sample hypothesis test in velocity space, which neither requires binning of the data, nor any fitting of parametrisations of the velocity distribution. We apply our method to realistic configurations of xenon and argon detectors such as XENONnT/DARWIN and DarkSide, and estimate the precision with which the DM mass can be determined. Once the dark matter mass is identified, the ratio of coupling strengths to neutrons and protons can be constrained by using the same data. The test can be applied for event samples of order 20 events, but promising sensitivities require $\gtrsim 100$ events.
\end{abstract}

%\keywords{Direct Detection}

\maketitle
%%%%%%%%%%%%%%%%%%%%%%%

\section{Introduction\label{sec:intro}}

So far dark matter (DM) has revealed its existence solely via gravitational interactions. We know its total energy density, both globally and locally, but neither its mass nor possible non-gravitational interactions are known. Motivated by the hypothesis that DM could be a WIMP (weakly interacting massive particle)~\cite{Goodman:1984dc}, there is a huge experimental effort in ton-scale noble gas direct detection (DD) experiments to search for the latter~\cite{Baudis:2014naa,Undagoitia:2015gya,Liu:2017drf}. Several projects will test in coming years large portions of well motivated WIMP parameter space before reaching the ultimate background induced by coherent neutrino scattering.

In order to extract the DM parameters from a positive DD signal,
assumptions about the local DM density and velocity distribution have
to be adopted. Typically a fit is performed assuming a Maxwellian DM
velocity distribution, denoted conventionally as the Standard Halo
Model (SHM). It is well known that in order to pin down the DM mass,
combining event spectra from experiments using different target nuclei
is beneficial, exploiting the different scattering kinematics, see
e.g.~Refs~\cite{Pato:2010zk,Peter:2013aha,Gluscevic:2015sqa}. However, both the local
energy density and the velocity distribution are subject to large
uncertainties, that translate directly into uncertainties regarding
the compatibility among different signals (and with upper limits) and
into the DM parameters.

In this paper we show how the DM mass can be determined
halo-independently from signals in two DD experiments by using a
nonparametric two-sample hypothesis test. The crucial observation is
that a value for the DM mass has to be adopted when transforming from
recoil energy to velocity. Therefore, the normalized weighted event
distributions in velocity space, which should be equal for both
experiments, will only be so for the true DM mass. This method can
also be applied to the case when DM--nucleus interactions are mediated
by a force carrier with mass $m_\phi$ in the 10--100~MeV range, where
both $m_\chi$ and $m_\phi$ have to be determined by the
data. Furthermore, we show how the relative coupling strength of DM to
protons and neutrons can be determined halo-independently from
the relative number of events in two DD experiments (properly weighted in velocity space).

The method presented here builds on and extends earlier work on halo-independent methods, first proposed in Refs.~\cite{Fox:2010bz, Fox:2010bu} and extensively used and extended to compare different experimental results, see e.g.\ Refs.~\cite{McCabe:2011sr, McCabe:2010zh, Frandsen:2011gi, HerreroGarcia:2011aa, HerreroGarcia:2012fu, Gondolo:2012rs, DelNobile:2013cta, DelNobile:2013cva, Bozorgnia:2013hsa, Frandsen:2013cna, Feldstein:2014gza, Fox:2014kua, Feldstein:2014ufa, Cherry:2014wia, Bozorgnia:2014gsa, Gelmini:2016pei, Gelmini:2017aqe, Ibarra:2017mzt, Kahlhoefer:2018knc, Catena:2018ywo}. The basic idea of these methods is that an integral of the velocity distribution, which can be extracted from the data, should be detector-independent. They have also been extended to compare with signals of neutrinos from the Sun~\cite{Blennow:2015oea,Ferrer:2015bta,Ibarra:2018yxq} and with collider signals~\cite{Blennow:2015gta,Herrero-Garcia:2015kga}.

Various versions of halo-independent methods to determine the DM mass have
been discussed previously \cite{Drees:2008bv, Feldstein:2014ufa,
  Feldstein:2014gza, Kavanagh:2012nr, Kavanagh:2013wba, Kavanagh:2013eya}, based either
on fitting moments of the velocity distribution or on fitting DM mass
and velocity distribution simultaneously. Our method is based on a
distribution-free statistical test, which does not require any binning
of data. It works for relatively small sample sizes, starting at
$\gtrsim 20$ events, while more robust results can be obtained for
$\gtrsim 100$ events. This is important in view of present bounds on
the scattering cross section \cite{Akerib:2016vxi, Cui:2017nnn,
  Aprile:2018dbl} which limits the possible number of DM scattering
events above the neutrino background.  The method is robust with
respect to energy reconstruction uncertainty and asymmetric event
numbers in the two experiments.

The paper is structured as follows. In Sec.~\ref{sec:rate} we introduce the relevant notation for DD event rates. We describe the halo-independent method to extract the DM mass in Sec.~\ref{sec:DMmass}, while in Sec.~\ref{sec:fnfp} we explain how to extract the ratio of couplings to protons and neutrons. In Sec.~\ref{sec:light} we show how the test can be applied in case of an interaction mediated by a light mediator. We give our conclusions in Sec.~\ref{sec:conclusions}.

%%%%%%%%%%%%%%%%%%%%%%%%%%%%%%%%%%%%%%%%%%%%%%%%%%%%%%%%%%%%%%%%%%
\section{The direct detection event rate} \label{sec:rate}
%%%%%%%%%%%%%%%%%%%%%%%%%%%%%%%%%%%%%%%%%%%%%%%%%%%%%%%%%%%%%%%%%%

For DM scattering elastically off single-target detectors with spin-independent (SI) interactions, the time-averaged differential event rate is given by
\begin{equation} \label{eq:R}
  \dfrac{d R_D}{d E_R}=\frac{\rho_\chi \bar\sigma A_{{\rm eff},D}^2}{2m_{\chi}\mu_p^2} F^2_D(E_R)\underbrace{\int_{v>v_{\text{m},D}}
    d^3v\, \frac{f\left(\vec{v}+\vec{v}_e\right)}{v}}_{\equiv \eta(v_{\text{m},D})}\,,
\end{equation}
where by kinematics
\begin{equation}\label{eq:vm}
 v_{\text{m},D} =  \sqrt{\frac{m_{A_D} E_R}{2\,\mu_{A_D}^2} }
\end{equation}
is the minimum DM velocity that detector $D$ is sensitive
to for a given recoil energy $E_R$. Following the notation of
Ref.~\cite{Schwetz:2011xm}, we define an effective target mass number
by
\begin{equation}\label{eq:Aeff}
A_{{\rm eff},D}^2 \equiv 2 [Z_D\cos\theta+(A_D-Z_D)\sin\theta]^2 \,,
\end{equation}
with $\tan\theta\equiv f_n/f_p$ being the ratio of couplings to
neutrons $f_n$ and protons $f_p$, and $\bar\sigma =
(\sigma_p+\sigma_n)/2$ the zero-momentum transfer cross section for
DM--nucleon scattering averaged over scattering on protons and
neutrons. If several isotopes of an element are present in the
detector, Eq.~\eqref{eq:Aeff} has to be replaced by a sum over the
isotopes, weighted by the relative abundance~\cite{Schwetz:2011xm}. Furthermore,
$m_{A_D}$ is the mass of the target nucleus in experiment $D$,
$\mu_{p}\, (\mu_{A_D})$ is the proton (nucleus) reduced mass, and
$F_D(E_{R})$ is the SI nuclear form factor.  $f(\vec v)$ describes the
distribution of DM particle velocities in the galaxy rest frame, and
$\rho_\chi$ is the DM local energy density. For our simulations, we
adopt the Helm parametrisation for $F_D(E_{R})$,
and we use the SHM, i.e., a Maxwellian velocity distribution, cut-off at the galactic escape velocity $v_{\rm esc}=544\,{\rm km/s}$, and with local energy density
$\rho_\chi=0.4\,{\rm GeV\, cm^{-3}}$~\cite{McCabe:2010zh}. We neglect
the time-dependent velocity of the Earth around the Sun.

The total number of events in detector $D$ above the energy threshold $E_{{\rm th},D}$ is given by 
\begin{equation}\label{eq:events}
n_D= M_{D} T_D \int_{E_{{\rm th},D}} d E_R\, \dfrac{d R_D}{d E_R}\,,
\end{equation}
where $M_{D} T_D$ is the exposure (detector mass times measurement
time). We ignore here possible detector- and energy-dependent
resolution and efficiency functions. We comment later on their impact.

\begin{table}
  \begin{tabular}{lccccc}
    \hline\hline Xe/Ar
    & $M_DT_D$ [t yr] & $E_{\rm th}$ [keV] &
    \multicolumn{3}{c}{\# of events} \\
    $m_\chi$ [GeV] & & & 20 & 50 & 100\\
    \hline
    \emph{Conservative} & 10/20 & 5/10 &  52/23 & 117/45 & 89/38 \\
    \emph{Optimistic} & 40/40 & 3/8 & 357/59 & 569/99 & 410/83\\
    \hline\hline
  \end{tabular}
  \caption{Parameters of our default configurations for the
    xenon/argon DM experiments denoted as \emph{conservative} and \emph{optimistic}.  We give the assumed exposure (detector
    mass $M_D$ times measurement time $T_D$) and the energy
    thresholds. The last three columns show the expected number
    of events for DM masses of 20, 50, and 100~GeV, assuming a
    fiducial SI cross section of $\bar\sigma = 5\times 10^{-47}$~cm$^2$ generated by a heavy mediator, and equal couplings to neutrons and
    protons. \label{tab:experiments}}
\end{table}

For our numerical calculations we simulate realistic realizations of
xenon and argon experiments, as several experiments using these
targets are planned: LZ~\cite{Akerib:2015cja},
PandaX~\cite{Cui:2017nnn}, XENONnT~\cite{Aprile:2014zvw}, and
ultimately DARWIN~\cite{Aalbers:2016jon}, using xenon;
DEAP-3600~\cite{Fatemighomi:2016ree}, ArDM~\cite{Calvo:2016hve},
DarkSide and Argo~\cite{Aalseth:2017fik}, using argon.  In
Tab.~\ref{tab:experiments} we define two benchmark configurations
which we denote as \emph{conservative} and \emph{optimistic}.  For comparison, XENONnT
and LZ plan for detector masses of order 10~t, while the goal of
DARWIN is 40~t of xenon. The DarkSide collaboration envisages a 20~t (100~t)
argon detector mass for DarkSide-20k (Argo). The event numbers quoted in the table are
obtained for a fiducial cross section of $\bar\sigma = 5\times
10^{-47}$~cm$^2$, close to the current XENON1T
limit~\cite{Aprile:2018dbl}.  For a zero-background experiment event
numbers are proportional to the product $M_DT_D \bar\sigma$.  The
assumed energy thresholds are motivated by those achieved by
the currently running experiments, as well as the target numbers quoted
in the respective proposals, taking into account the zero-background
assumption. In our numerical analysis, we generate many instances
(typically $10^3$) of recoil energy event distributions for the two
experiments by Monte Carlo, with the total number of events drawn from
a Poisson distribution with mean given by Eq.~\eqref{eq:events}.

%%%%%%%%%%%%%%%%%%%%%%%%%%%%%%%%%%%%%%%%%%%%%%%%%%%%%%%%%%%%%%%%%%
\section{Extracting the dark matter mass\label{sec:DMmass}}
%%%%%%%%%%%%%%%%%%%%%%%%%%%%%%%%%%%%%%%%%%%%%%%%%%%%%%%%%%%%%%%%%%

The key observation used in halo-independent methods is that, while
the prefactor in Eq.~\eqref{eq:R} depends on the target nucleus,
$\eta(v_{\rm m})$ is a detector-independent quantity~\cite{Fox:2010bz,
  Fox:2010bu}. Therefore, if two signals are observed in detectors
using different targets, one can compare their properly weighted
distributions in the overlapping velocity space for different DM
masses, and they will only agree for the true DM mass. In the following we will employ a non-parametrical two-sample hypothesis test
to infer the correct DM mass from two DD event samples.

Suppose detectors 1 and 2 observe $n_1$ and $n_2$ DM induced events,
respectively, with certain recoil energies $E_R^{i,D}$, where
$i=1,...,n_D$, $D=1,2$. For an assumed value of the DM mass $m_\chi$,
these recoil energies can be transformed into velocities via
Eq.~\eqref{eq:vm}. It is more convenient to work with the square of
the minimal velocity, since then the Jacobian of the transformation is
just a constant. Therefore we define the transformed event samples
$x_i = v_{i,1}^2(m_\chi)$ and $y_j = v_{j,2}^2(m_\chi)$.  If the true
value for $m_\chi$ has been used in this transformation, the random
variables $x_i$ and $y_j$ will be distributed according to probability
distribution functions (PDF) proportional to $F^2_1 (v^2) \eta(v^2)$
and $F^2_2 (v^2) \eta(v^2)$, respectively. We can now weigh the
distribution of each experiment with the form factor of the other
experiment. We define
\begin{align}
  h(v^2) &\equiv F^2_1 (v^2) F^2_2 (v^2) \eta(v^2) \,,\quad
  \label{eq:h} \\
  \tilde h(v^2) &\equiv {\mathcal N}\, h(v^2) 
  \quad\text{with}\quad
  \int_{v^2_{\rm m,th}} dv^2 \tilde h(v^2) = 1 \,,
  \label{eq:htilde}
\end{align}
where ${\mathcal N}$ is a normalization constant and in the arguments of the form factors $F_D^2(v^2)$ velocity is converted into energy using the relation corresponding to the experiment $D$, see Eq.~\eqref{eq:vm}. By construction the PDF $\tilde h(v^2)$ will be identical for the two event
samples, if the correct DM mass is used to convert recoil energy
into squared-velocities. 

In order to apply this idea to the data samples $x_i$ and $y_j$ we
consider the corresponding cumulative distribution function (CDF),
$H(v^2) = \int_{v^2_{\rm m,th}}^{v^2} dx\, \tilde h(x)$. It can be
estimated from the two data samples in the following way:
\begin{align}
  \hat{H}_{(1)}(x) &=\frac{1}{\omega_{\rm t, 1}} \sum_{i=1}^{n_1} 1_{x_i \le x}\, \omega_{i,1}\,,
  \nonumber \\
  \hat{H}_{(2)}(x) &=\frac{1}{\omega_{\rm t, 2}} \sum_{j=1}^{n_2} 1_{y_j \le x}\, \omega_{j,2}\,,
  \label{eq:ecdf}
\end{align}
where $1_{x_i \le x}$ is equal to 1 for $x_i \le x$, and zero
otherwise, and
\begin{equation}\label{eq:weights}
\omega_{i,1} \equiv F^2_2[E_{R,2}(v^2_{i,1})]\,,\quad
\omega_{\rm t, 1} \equiv \sum_{i=1}^{n_1} \omega_{i,1}
\end{equation}
and similar for the weights $\omega_{j,2}$ of the events $y_j$ of the
second sample.\footnote{The Radon-Nikodym theorem guarantees the
  convergence of the weighted empirical distribution, as long as the
  weighted distribution is equal or smaller than the original
  distribution of the data.  This is the reason why we weigh the
  events of experiment~1 by the form factor of experiment~2, and
  viceversa, using that $F_D^2(E_R) \le 1$.}  As we normalize by the
sum of the weights, the prefactors and the Jacobian of the
transformation to velocity space drop out. The lower integration
boundary $v^2_{\rm m,th}$ in Eq.~\eqref{eq:htilde} is determined by
the energy thresholds of the two detectors: for a given DM mass
$m_\chi$ the two recoil energy thresholds are transformed into
velocity-squared using Eq.~\eqref{eq:vm} and $v^2_{\rm m,th}$ is
chosen as the larger of the two. The sums in Eqs.~\eqref{eq:ecdf} and
\eqref{eq:weights} include only events with $x_i,y_j > v^2_{\rm
  m,th}$.  This ensures that only events which probe overlapping
regions in $v^2$-space are considered.\footnote{The upper analysis
  limits are assumed to be high enough, such that they never play a
  role.} Note that in certain cases there will be no events in the
overlapping region. In these cases our test cannot be applied.

Several standard statistical tools are available to test whether two
empirical CDFs emerge from the same underlying PDF (which is the null hypothesis in the following), for instance the
Kolmogorov-Smirnov, Cram\'er-von Mises, or Anderson-Darling tests, see
e.g., Ref.~\cite{monahan}. In the following we will present results
based on the Cram\'er-von Mises (CvM) test, which for our application
shows somewhat better statistical properties than alternative tests.
The corresponding test statistic $T_{\rm CvM}$ is defined as
\begin{align} 
  \frac{(\tilde n_1+\tilde n_2)^2}{\tilde n_1 \tilde n_2}T_{\rm CvM} &=
  \sum_{i=1}^{n_1}\left[\hat{H}_{(1)}(x_i) - \hat{H}_{(2)}(x_i)\right]^2 \nonumber\\ 
  &+ \sum_{j=1}^{n_2}\left[\hat{H}_{(1)}(y_j) - \hat{H}_{(2)}(y_j)\right]^2 \,,
  \label{eq:TCvM}
\end{align}
and it has a known asymptotic distribution under the null
hypothesis~\cite{10.2307/2236446}, which can be used to calculate a
$p$-value for a given observed value $T_{\rm CvM}^{\rm obs}$, i.e.,
the probability of obtaining $T_{\rm CvM} > T_{\rm CvM}^{\rm obs}$.
The effective event numbers $\tilde n_D \equiv (\sum
\omega_{i,D})^2/\sum\omega_{i,D}^2$ in Eq.~\eqref{eq:TCvM} take into account that the actual
events have been drawn from the unweighted
distribution~\cite{monahan}. The null hypothesis (i.e., the samples
emerge from the same PDF) corresponds to the case that the correct DM
mass has been used to convert from recoil energy to velocity-squared
space.  Hence, in the plots below we will show the $p$-value as proxy
of the discriminating power for the DM mass: a low $p$-value will
indicate that the null-hypothesis has to be rejected at a certain
confidence level, and therefore that particular value of the DM mass
as well. We have checked, by explicit Monte Carlo simulations of DM
events, that under the null-hypothesis the test statistic defined as
in Eq.~\eqref{eq:TCvM} indeed follows the known distribution for the
CvM statistic, despite the re-weighting of the data necessary to cope
with the form factors.

The method can also serve to test if two signals are compatible with
each other under the DM hypothesis: if the $p$-value is small for any
DM mass, then one can conclude that at least one of the signals is not
consistent with the underlying DM hypothesis (e.g. elastic scattering,
spin-independent interactions). This is independent of whether the
couplings are isospin-conserving or violating (see also below).

\subsection{Numerical results \label{sec:results}}

\begin{figure*}[t!]\centering
	\includegraphics[width=0.325\linewidth]{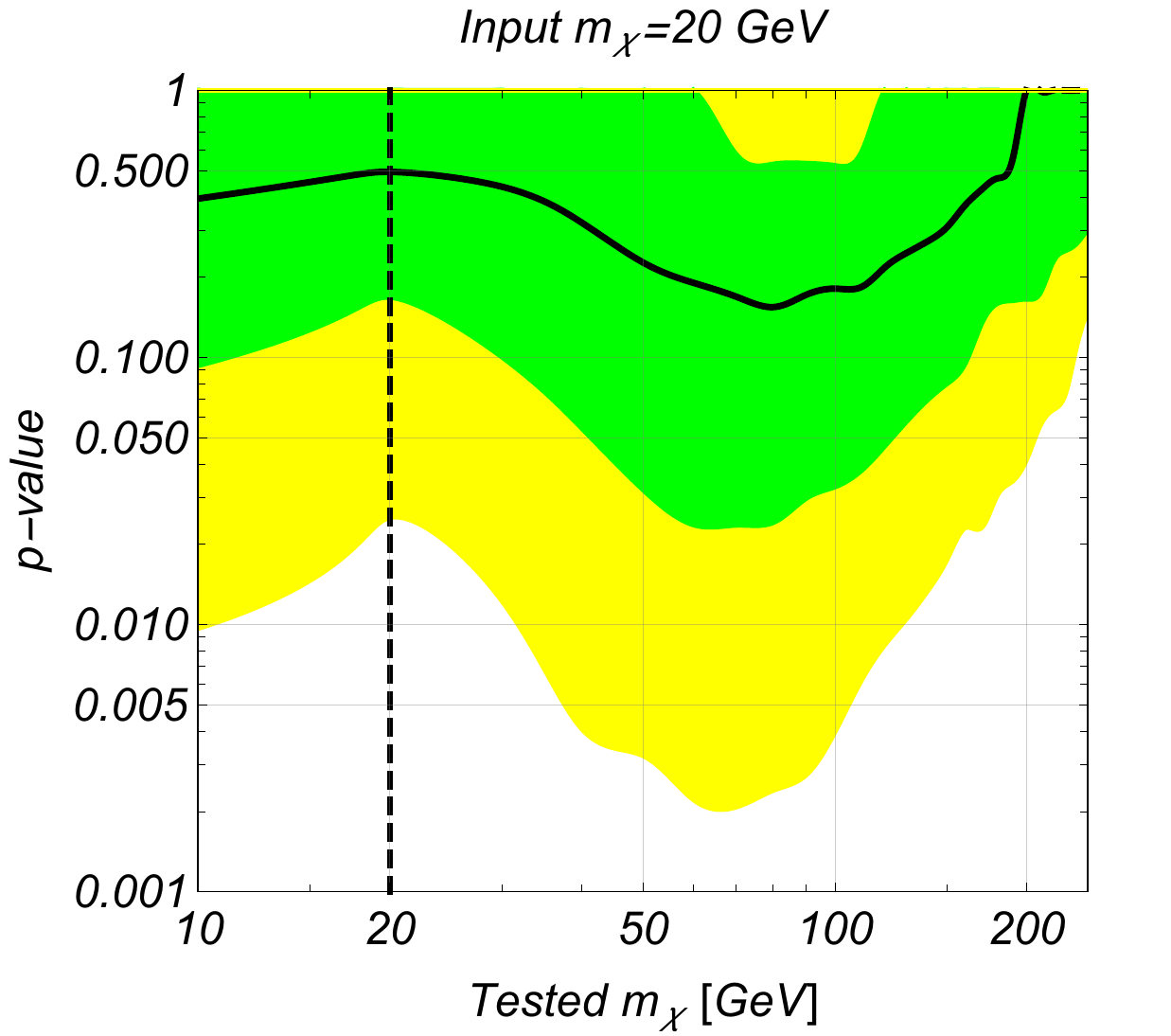}
	\includegraphics[width=0.325\linewidth]{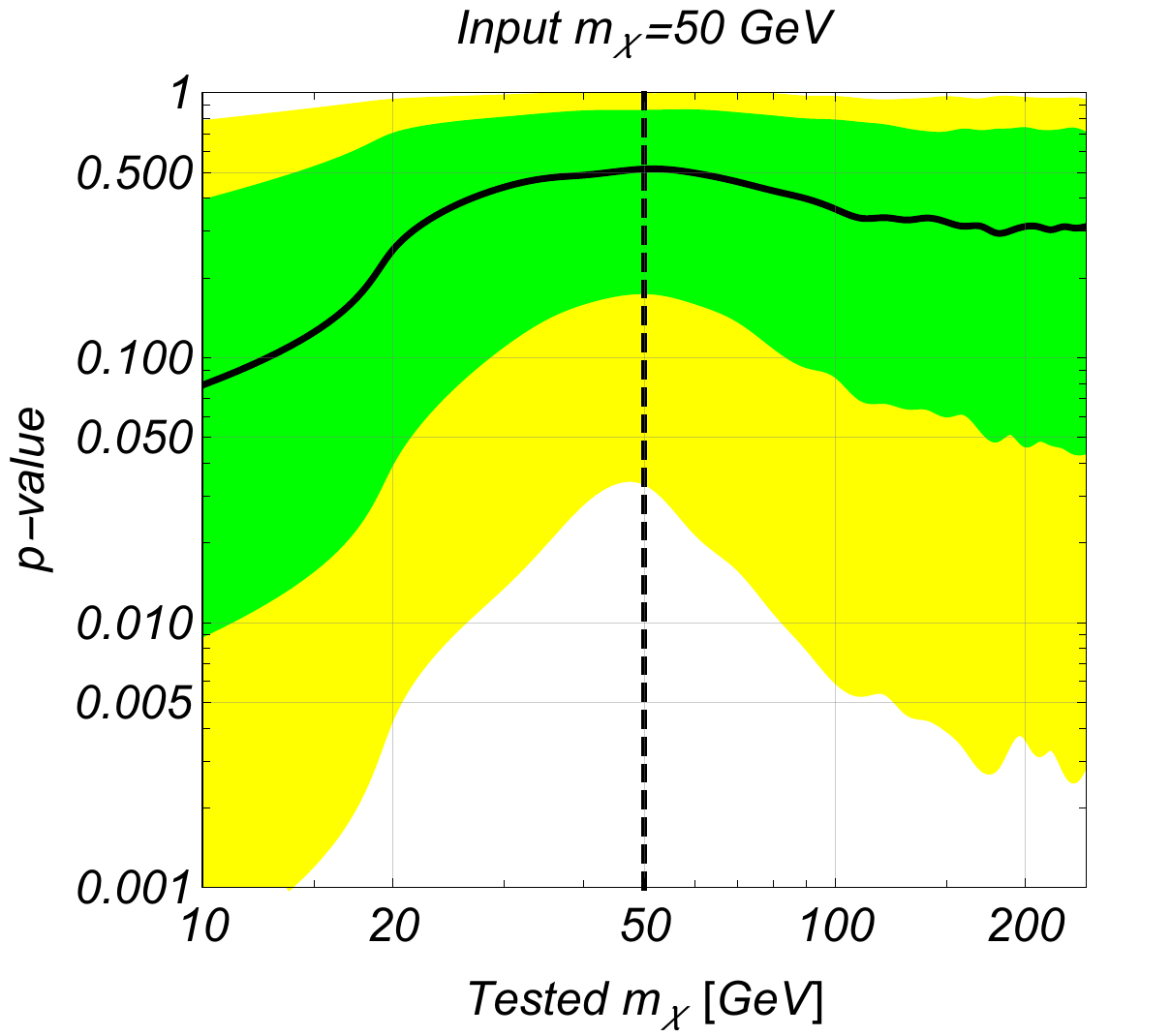}
	\includegraphics[width=0.325\linewidth]{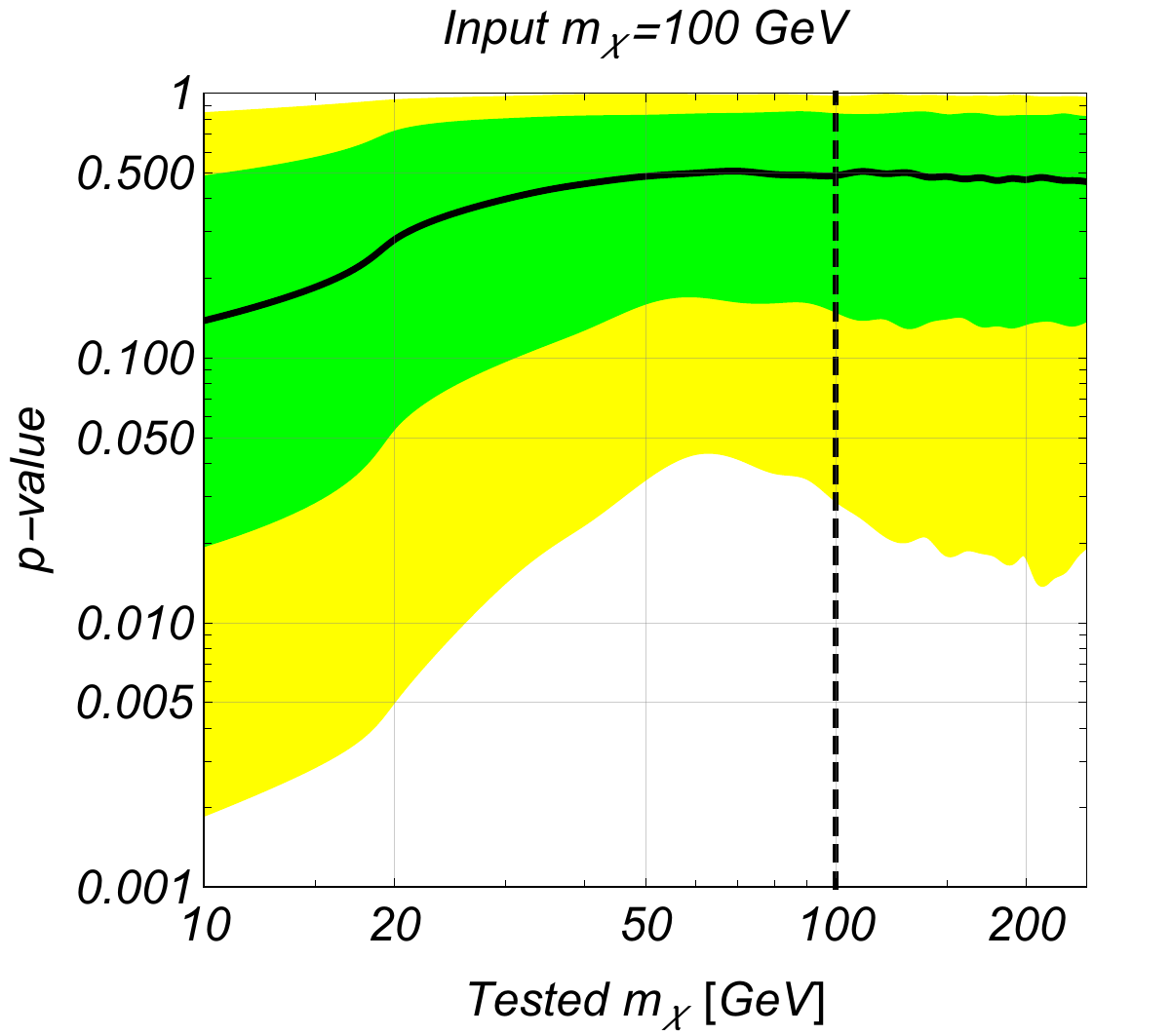}
	\caption{$p$-value for the tested dark matter mass, for
          different values of input DM mass shown as dashed vertical lines: $20$ GeV (left), $50$ GeV
          (middle) and $100$ GeV (right). The \emph{conservative}
          configuration has been assumed (see
          Tab.~\ref{tab:experiments}). We have generated $10^3$
          random data samples. The black curve corresponds to the
          median $p$-value, whereas the shaded green and yellow regions indicate the
          range of $p$-values obtained in 68\% and 95\% of the
          cases, respectively.} \label{fig:rej_rat_cons}
\end{figure*}

\begin{figure*}[t!]\centering
	\includegraphics[width=0.325\linewidth]{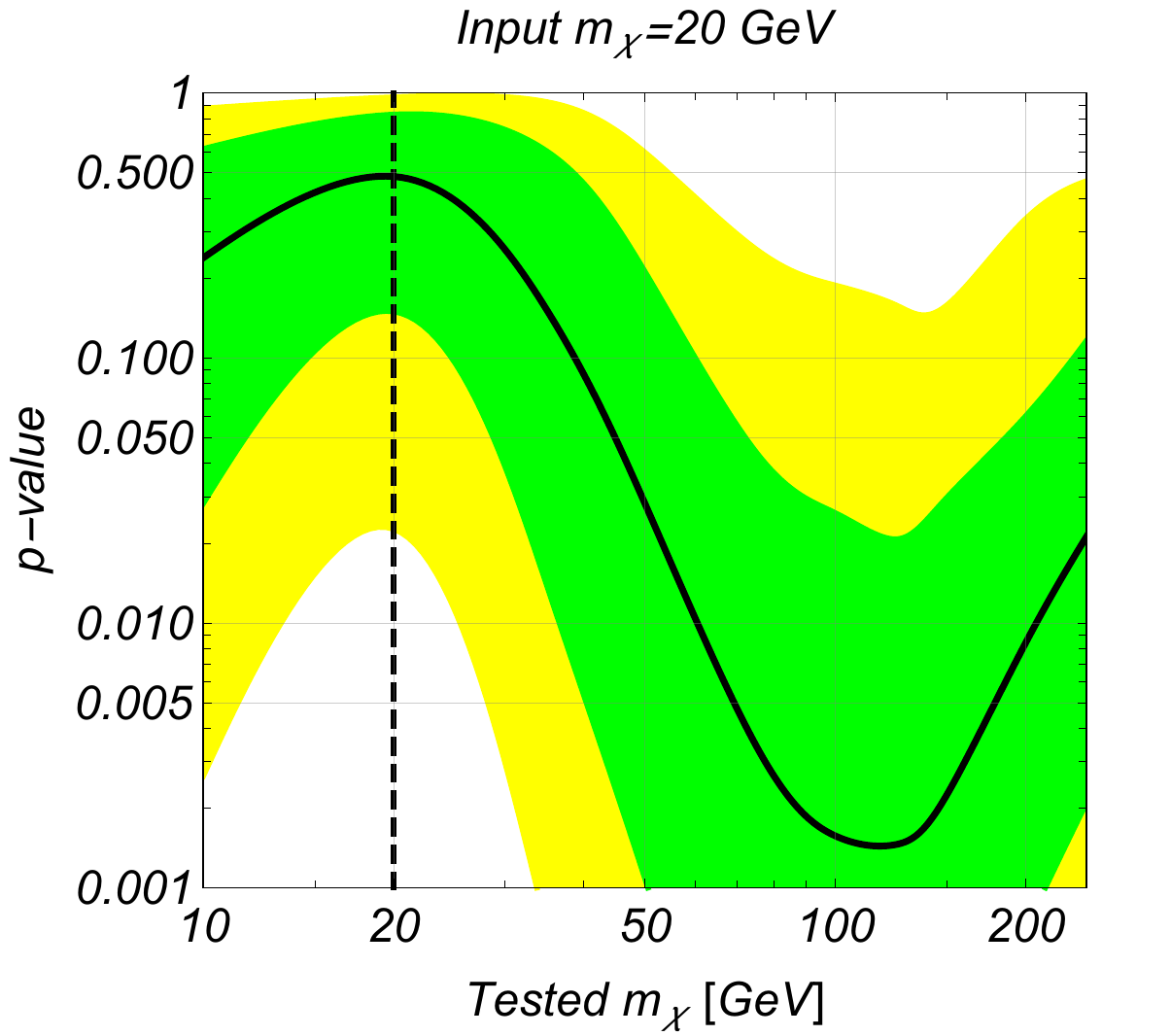}
	\includegraphics[width=0.325\linewidth]{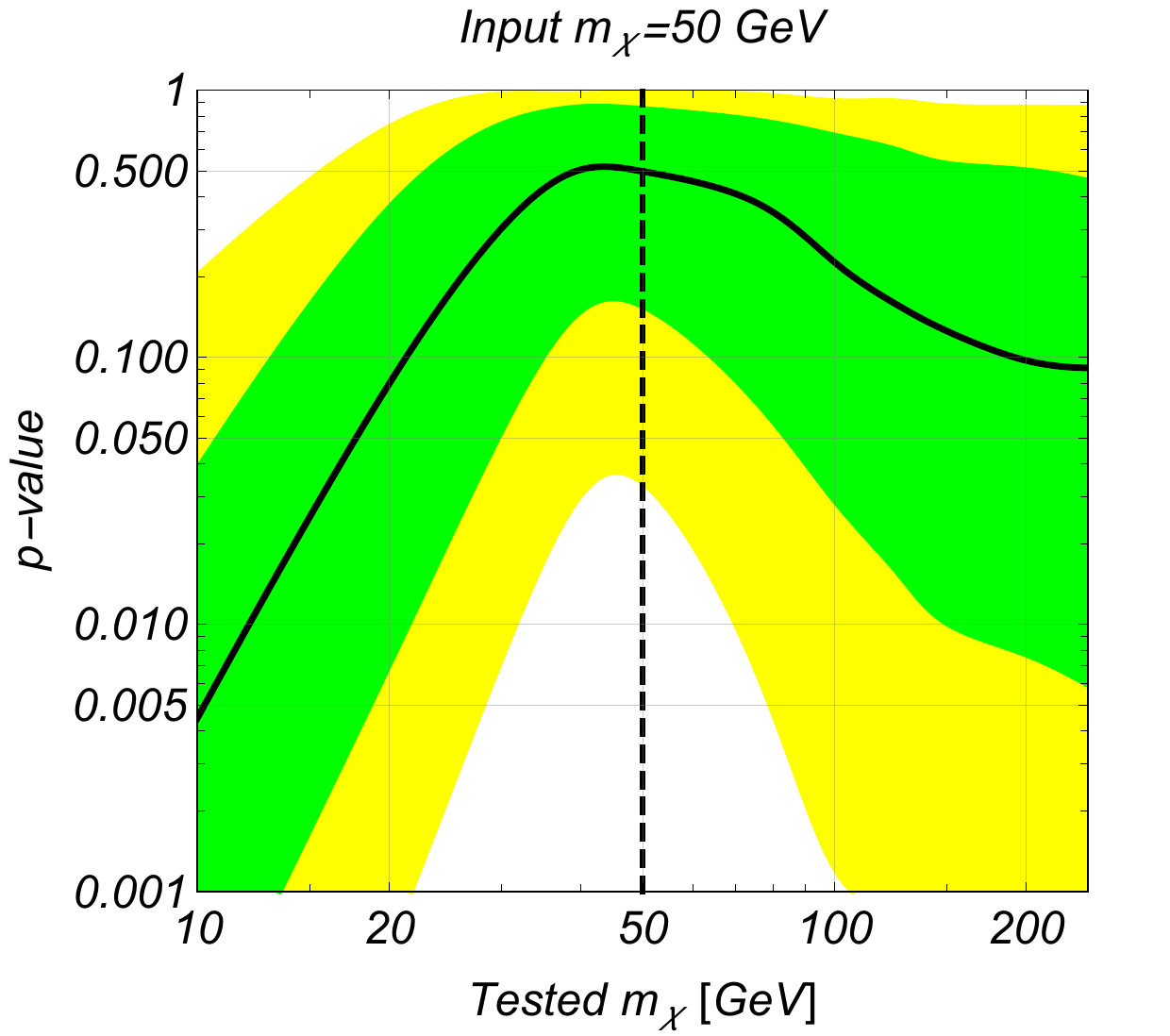}
	\includegraphics[width=0.325\linewidth]{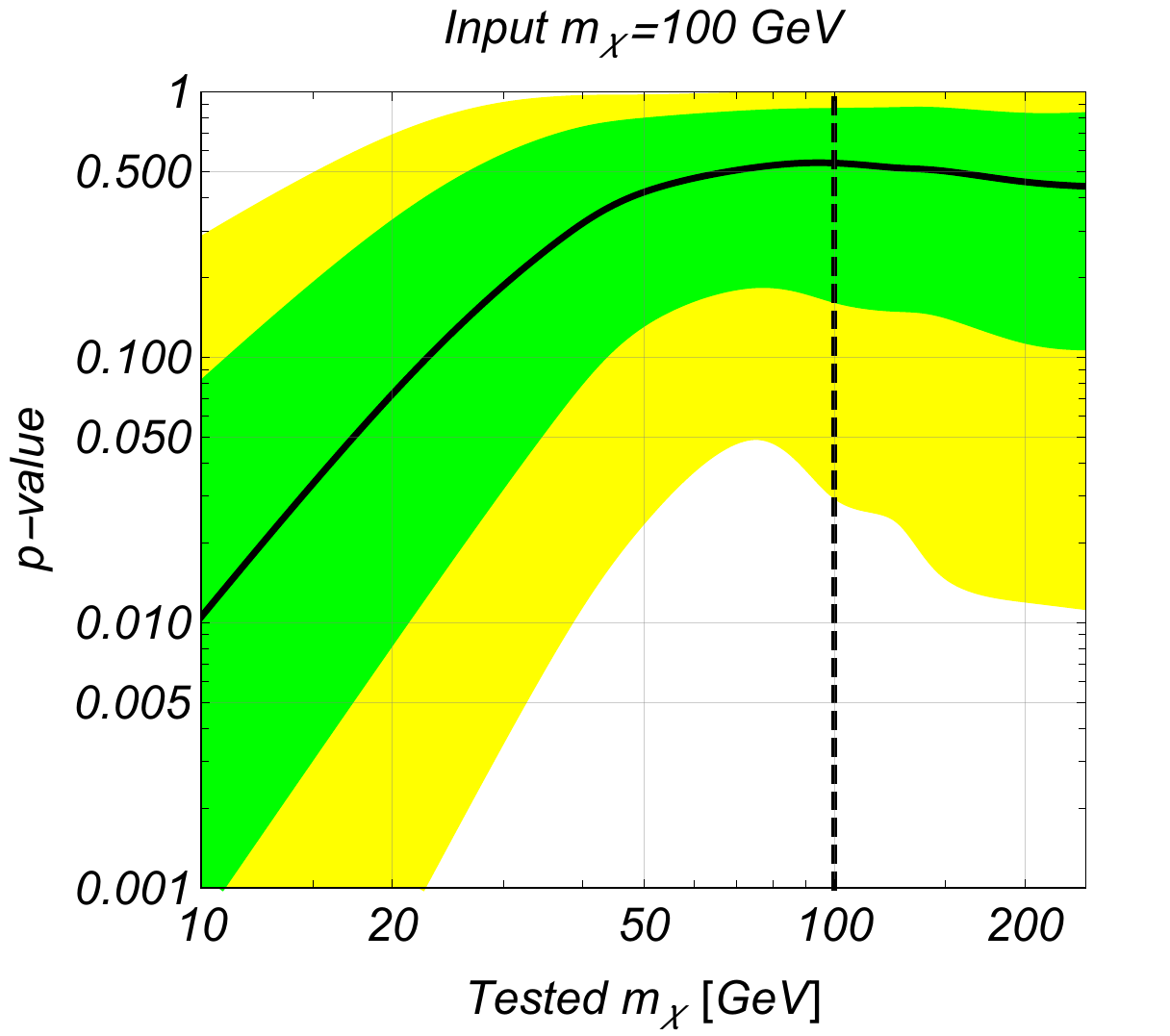}
	\caption{Same as Fig.~\ref{fig:rej_rat_cons} but assuming the
          \emph{optimistic} configuration (see Tab.~\ref{tab:experiments}).}
          \label{fig:rej_rat_opt}
\end{figure*}

In Figs.~\ref{fig:rej_rat_cons} and \ref{fig:rej_rat_opt} we show the
results of the numerical analysis for the \emph{conservative} and \emph{optimistic}
experimental configurations defined in Tab.~\ref{tab:experiments},
respectively.  We have generated $10^3$ random realizations of the
experiments, by assuming fixed true DM masses of 20, 50, and 100~GeV.
For a given true DM mass we then calculate the $p$-value of each
random data sample as a function of $m_\chi$ (denoted ``tested'' DM mass). The
plots show the median of the $p$-values (black curves), as well as the
range of $p$-values obtained in 68\% and 95\% of the cases (green and
yellow bands).

\begin{figure}
\centering
	\includegraphics[width=1\linewidth]{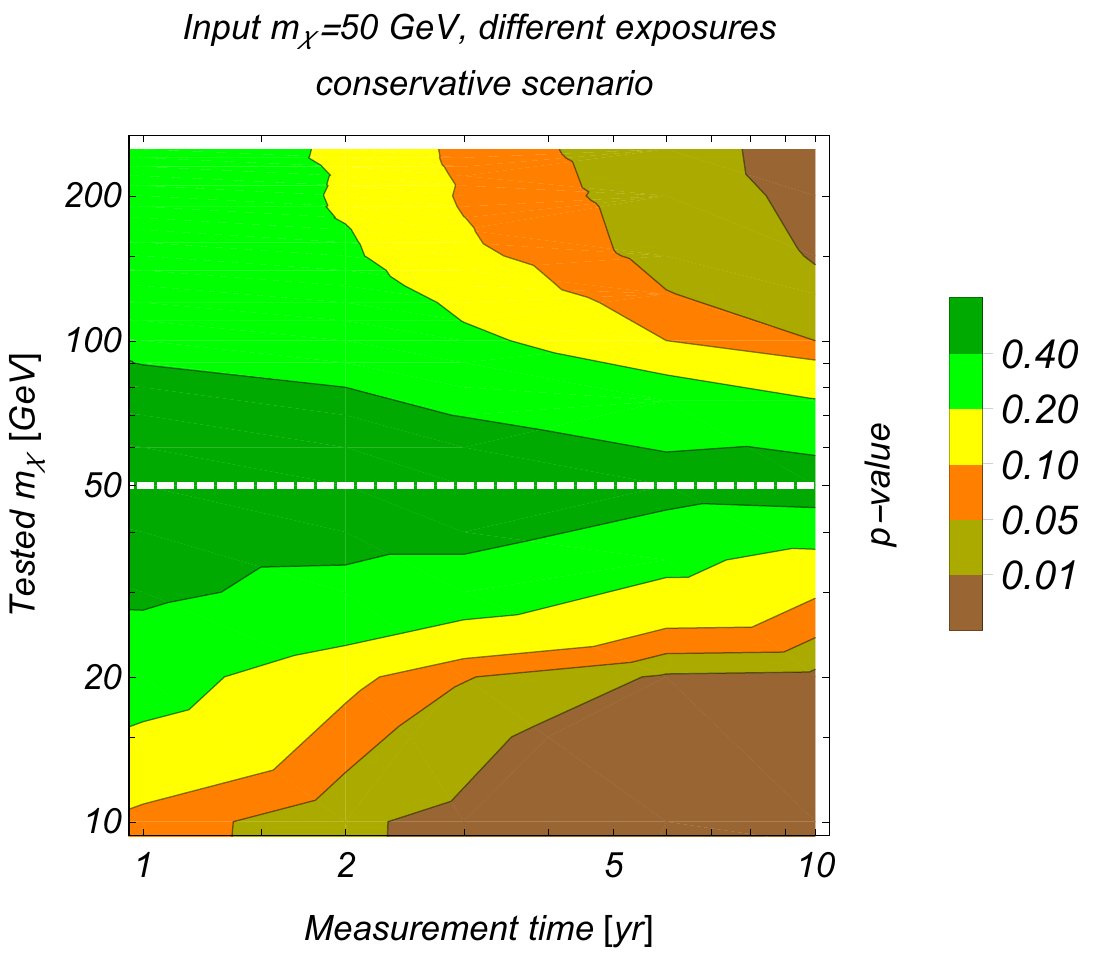}
	\caption{Median $p$-value of the reconstructed (tested) DM mass
          as a function of measurement time, for an assumed true DM
          mass of 50~GeV. We adopt the \emph{conservative} configuration from
          Tab.~\ref{tab:experiments} with detector masses of 10/20~t
          for xenon/argon.} 
          \label{fig:exposure}
\end{figure}

First, we observe that there is a rather large spread in the
$p$-values. E.g., while the median $p$-values for the \emph{conservative}
configuration are generally above 0.1, there is a rather high chance
that much stronger discrimination can be obtained, with $p$-values
even below 0.01, c.f.~Fig.~\ref{fig:rej_rat_cons}.  Conversely, even
for the \emph{optimistic} configuration chances are high that discrimination
against wrong DM masses is poor, c.f.~Fig.~\ref{fig:rej_rat_opt}.
Second, focusing on the median $p$-value, we see that for significant
DM mass determinations exposures similar to the \emph{optimistic} case may be
required, i.e., a few hundreds of events in xenon and around 100
events in argon.  From Fig.~\ref{fig:rej_rat_opt} we see that for the
average \emph{optimistic} configuration, DM masses of 20 and 50~GeV, can be
determined at the 90\%~CL to be in the ranges [7,\,38] and [21,\,190]~GeV,
respectively, and $m_\chi = 100$~GeV can be constrained to be $\ge
23$~GeV. In Fig.~\ref{fig:exposure} we show the uncertainty with which
a true DM mass of 50~GeV can be determined as a function of the
measurement time. We observe roughly a scaling with the square-root of
the exposure.\footnote{Note that we rescale here the \emph{conservative}
configuration. The \emph{optimistic} case has a different ratio of the
  detector masses in xenon and argon as well as different energy
  thresholds; therefore it cannot be obtained exactly by rescaling the
  \emph{conservative} configuration.} After a 10~year exposure for our
benchmark cross section and detector masses the range can be
constrained at 90\%~CL to [30,\,90]~GeV.

\begin{figure}
\centering
	\includegraphics[width=1\linewidth]{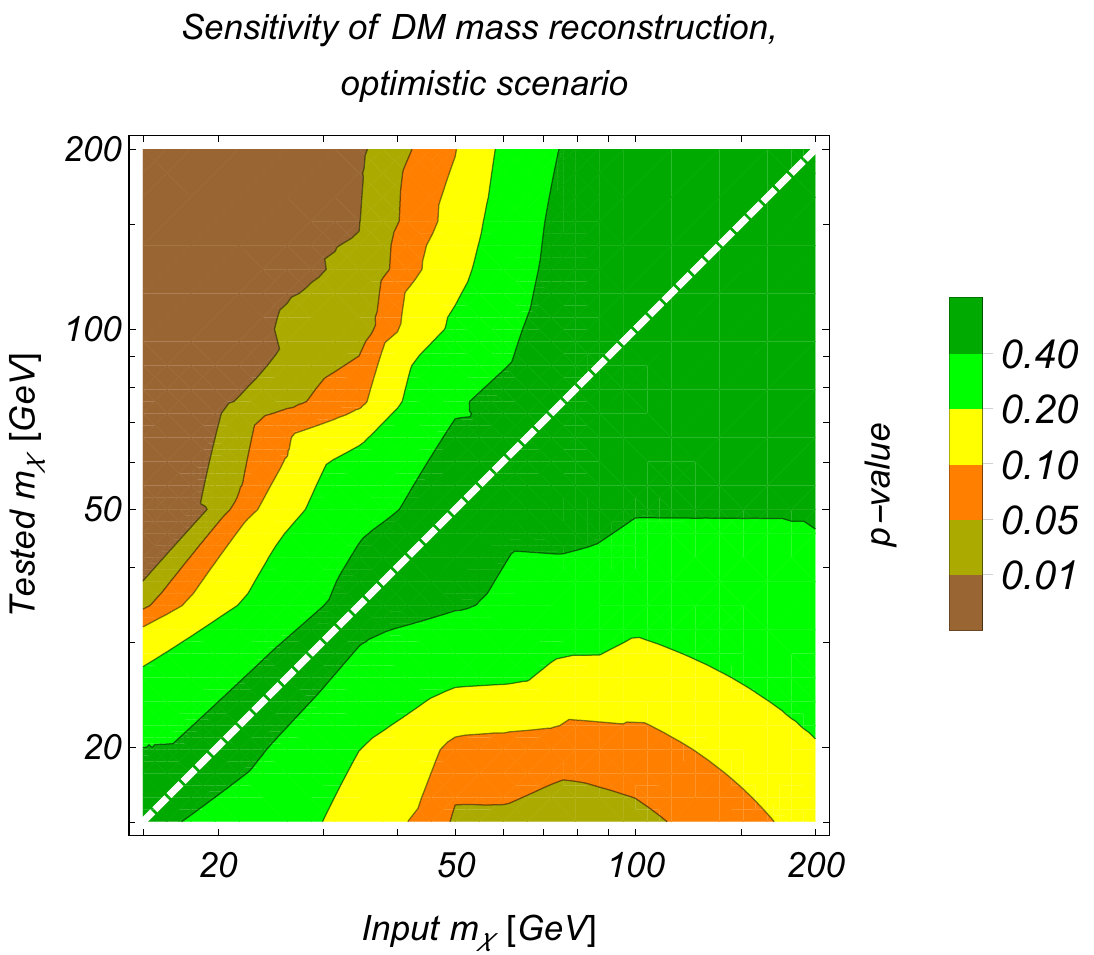}
	\caption{Contours of median $p$-value of the reconstructed
          (tested) DM mass as a function of the true (input) DM mass
          for the \emph{optimistic} configuration defined in
          Tab.~\ref{tab:experiments}.} \label{fig:mtr-mrec} 
\end{figure}

The precision with which $m_\chi$ can be determined as a function of
the true DM mass is shown in Fig.~\ref{fig:mtr-mrec} for the median
\emph{optimistic} configuration. We see that the test works fine for DM
masses in the range approximately from 20 to 70~GeV. For larger DM
masses only a lower bound can be obtained.  This is to be expected,
since for $m_\chi \gg m_A$, $v_{\rm m}$ and therefore $\eta(v_{\rm
  m})$ become independent of the DM mass, c.f.~Eq.~\eqref{eq:vm}. This
behaviour is not specific to our test; it follows from general
kinematics and \emph{any} DM determination from DM--nucleus scattering
has this property.  Note that for calculating Fig.~\ref{fig:mtr-mrec}
we fix $M_DT_D \bar\sigma$ to the value defined in
Tab.~\ref{tab:experiments}. This implies that the event rate decreases
linearly with $m_\chi$ for $m_\chi \gg m_A$, see Eqs.~\eqref{eq:R} and
\eqref{eq:events}. This explains the decrease of the lower bound on
$m_\chi$ for input values $\gtrsim 100$~GeV in
Fig.~\ref{fig:mtr-mrec}. We have checked that if event numbers are
kept constant when changing the true $m_\chi$ the lower bound remains
approximately constant.

Let us comment on the increase of the $p$-value visible for large DM
masses in the case of true $m_\chi = 20$~GeV (left panels in
Figs.~\ref{fig:rej_rat_cons} and \ref{fig:rej_rat_opt}). In this
region it may happen, that the events of the two experiments fall into
distinct regions in $v_{\rm m}$ space, i.e., there is no overlapping
range of $v_{\rm m}$ values. The exact region in which this occurs is,
to some extent, a consequence of the SHM assumption and the escape
velocity used in generating our events.  In such cases the test cannot
be applied and formally the test statistic is zero, indicating that
data is consistent with the null-hypothesis, i.e., such values of the
DM mass can be consistent with the data. In such a case other
diagnostic tools have to be employed, to find out whether data are
consistent. For instance, one could check whether the situation of
non-overlapping events in $v_{\rm m}$-space is consistent with the
fact that $\eta(v_{\rm m})$ has to be a decreasing function (which would
require an additional assumption on the ratio of couplings to neutrons
and protons, though).

\subsection{Robustness against energy resolution and background\label{sec:robustness}}

In the previous analysis, we have assumed perfect energy resolution
and efficiency and zero background. In order to study whether our
method is robust also in less ideal situations we have introduced a
constant gaussian energy resolution when generating the Monte Carlo
data, but still assume perfect resolution when applying the DM mass
test. We find that for energy resolutions below $2$~keV the test
results are essentially unmodified. Similarly, we also simulated the
case of constant or exponential backgrounds in the data, but ignoring
it when applying the test. The discriminating power of our method is
found to be unaffected as long as the background is below roughly
$10\%$ of the DM signal. Both results illustrate that indeed the
method can be realistically applied when two signals are observed.

We have also checked that the test performs slightly better for
similar number of events in both experiments (therefore it is
desirable to have a larger exposure for argon than for xenon, for
example). Note however, that our benchmark scenarios defined in
Tab.~\ref{tab:experiments} have rather asymmetric event numbers, and
therefore our test works also fine if one of the two experiments has less
events than the other.

%%%%%%%%%%%%%%%%%%%%%%%%%%%%%%%%%%%%%%%%%%%%%%%%%%%%%%%%%%%%%%%%%%%%%%%%%
\section{Ratio of couplings to neutrons and protons\label{sec:fnfp}}
%%%%%%%%%%%%%%%%%%%%%%%%%%%%%%%%%%%%%%%%%%%%%%%%%%%%%%%%%%%%%%%%%%%%%%%%%

Let us assume that the DM mass can be determined with sufficient
precision. Then it is possible to use the data from the two
experiments to constrain also the ratio of couplings to neutrons and
protons, i.e., $\theta=\arctan{(f_n/f_p)}$, see Eq.~\eqref{eq:Aeff}.
For simplicity we focus on the heavy mediator case, for the generalization
to light mediators see section~\ref{sec:light}, and in particular footnote~\ref{foot:lm}.  Let us consider the
following quantities:
\begin{equation}
  q_D \equiv M_{D}T_D \frac{\rho_\chi \bar\sigma }{m_{\chi}\, m_{A_D}}\frac{\mu_{A_D}^2}{\mu_p^2} A_{{\rm eff},D}^2 \int_{v^2_{{\rm m,th}}} dv^2 h(v^2) \,,
\end{equation}
with $h(v^2)$ defined in Eq.~\eqref{eq:h}, $v^2_{{\rm m,th}}$
determined as described above, and where we have taken into account the
Jacobian from changing integration variables from $E_R$ to $v^2$. We
see that the ratio
\begin{equation}\label{eq:Aeff-ratio}
  \frac{q_1 m_{A_1}/(M_1T_1\mu_{A_1}^2)}{q_2 m_{A_2}/(M_2T_2\mu_{A_2}^2)} =
  \frac{A_{\rm eff,1}^2(\theta)}{A_{\rm eff,2}^2(\theta)} 
\end{equation}
only depends on $\theta$ and is independent of the halo integral as
well as global factors such as the total cross section and the local
DM density. The quantity $q_D$ can be estimated from data. Indeed, it
corresponds to the total weights defined in Eq.~\eqref{eq:weights},
which are obtained by evaluating the form factor of the other
experiment at the recoil energies of the observed events:
\begin{equation}
\hat q_D = \omega_{\rm t,D} \,.  
\end{equation}

\begin{figure}
\centering
	\includegraphics[width=0.9\linewidth]{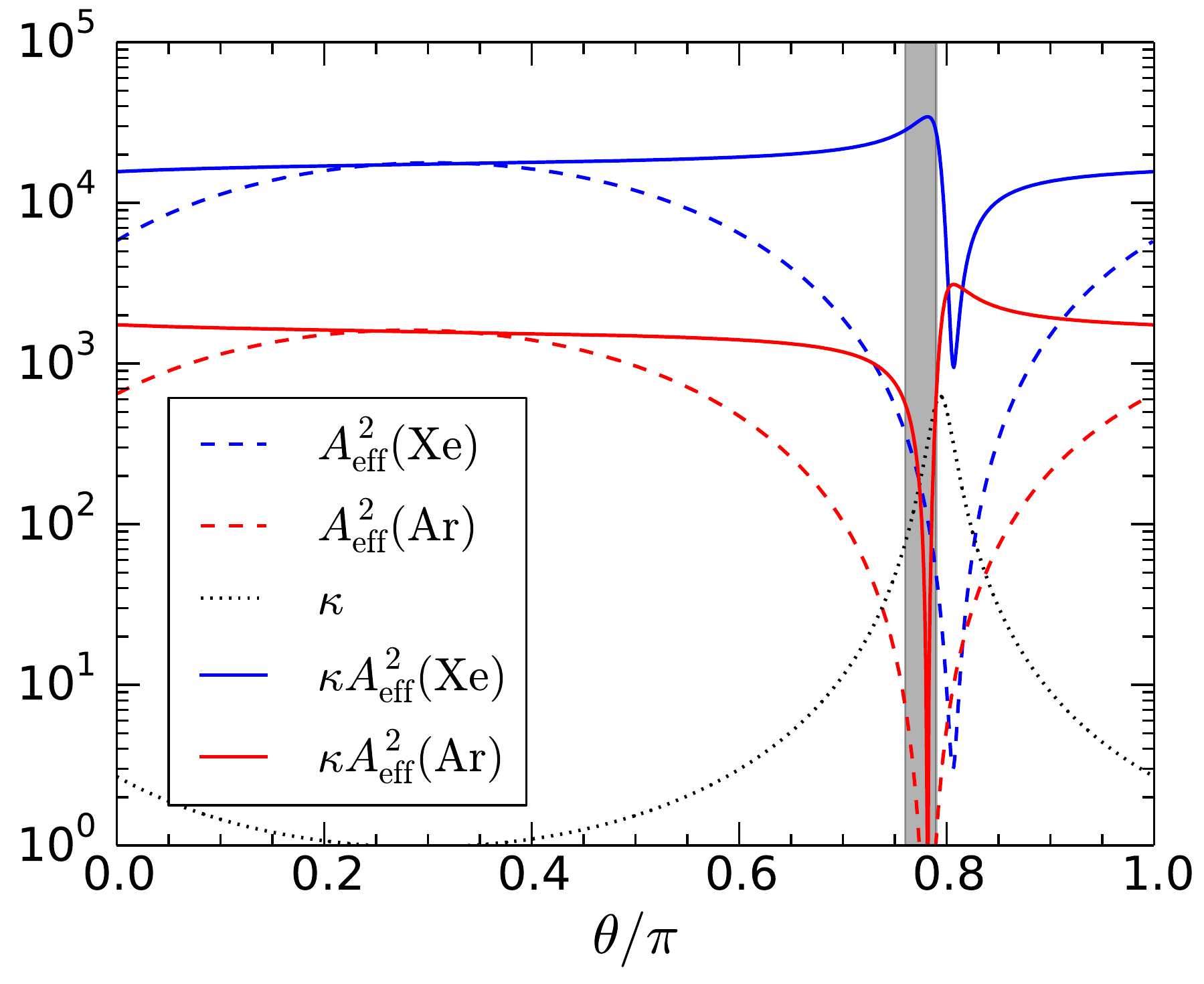}
	\caption{Effective mass numbers-squared (dashed) and rescaled
          ones (solid) for xenon (blue) and argon (red), and the
          applied rescale factor (dotted) are shown as a function of
          $\theta = {\rm arctan}(f_n/f_p)$, with $f_n$ and $f_p$ being
          the coupling to neutron and proton, respectively. The shaded
          region indicates the range where the event rate in argon
          detectors is strongly suppressed.}\label{fig:Aeff}
\end{figure}

In Fig.~\ref{fig:Aeff} we show the effective mass number squared for
xenon and argon (dashed curves). Event numbers are strongly suppressed
if $\tan\theta \approx -Z/(A-Z)$. We see from the plot that this
cancellation happens for $\theta \approx 0.8\pi$ for both
elements. Argon detectors employ depleted argon which consists basically
only of $^{40}$Ar. Therefore the cancellation can be complete. For xenon
we take into account the natural isotope composition, and therefore
the cancellation is never exact. In order to maintain events for at
least one detector, we rescale the effective cross section by an
arbitrary factor $\kappa(\theta)$, shown as dotted curve in the plot. It has
been chosen in such a way that the number of events remain approximately constant also
close to the cancellation region, at least for one of the two
experiments, see solid curves in Fig.~\ref{fig:Aeff}. Obviously, if
one of the experiments does not see events, the DM mass cannot be
determined, and therefore our method cannot be applied. This case is
indicated by the shaded region, where the event rate in the argon experiment is strongly
suppressed. Note that due to the isotope
distribution in xenon, we can have sizeable event numbers for xenon
even close to the cancellation region.

\begin{figure}
\centering
	\includegraphics[width=0.9\linewidth]{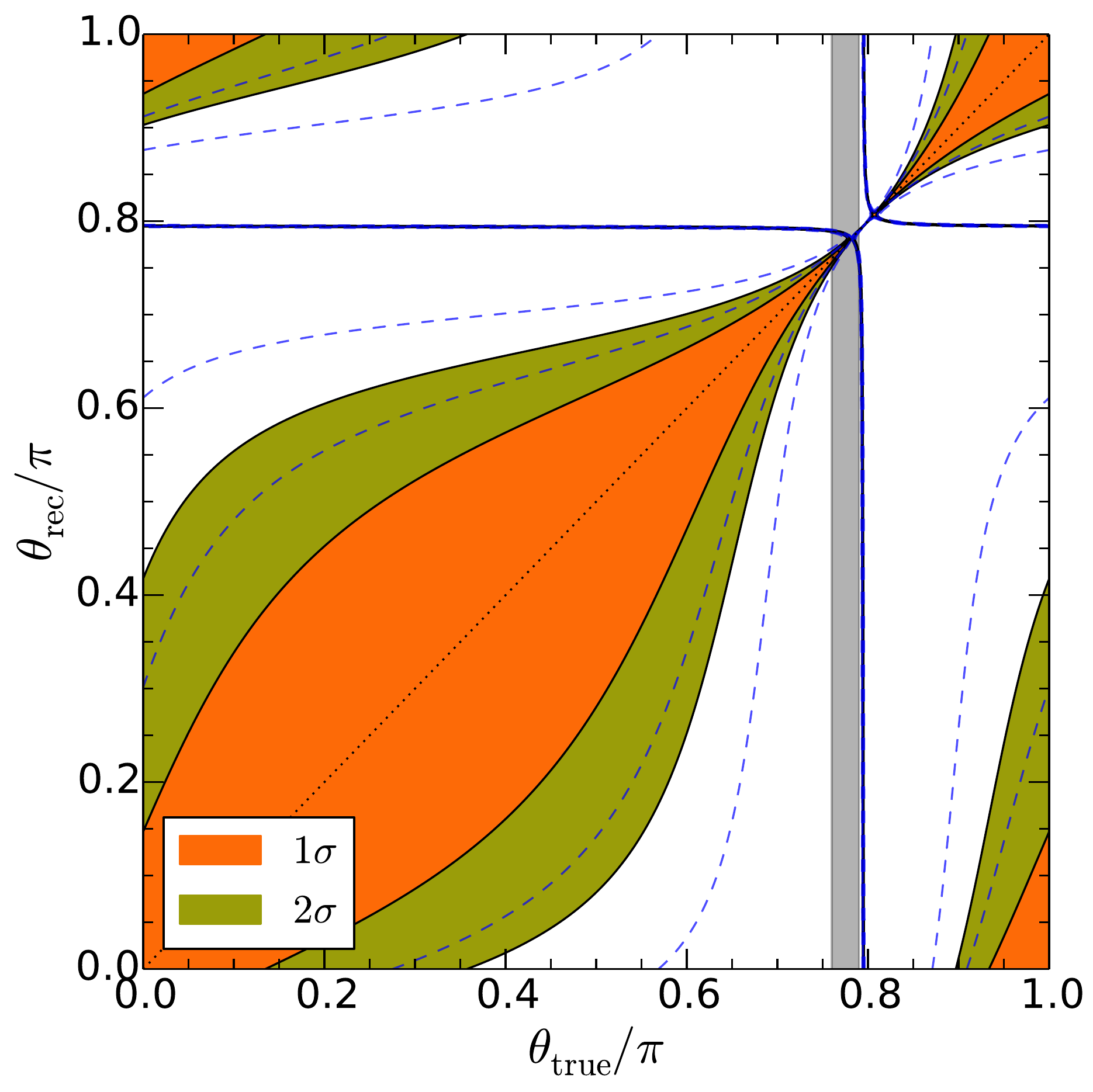}
	\caption{Reconstructed relative coupling to neutrons and
          protons as a function of the true value for the \emph{optimistic}
          experimental configuration with re-scaled event numbers as
          shown in Fig.~\ref{fig:Aeff}.  For a given $\theta_{\rm
            true}$, the interval in $\theta_{\rm rec}$ corresponding
          to the white region can be excluded at $2\sigma$.  Dashed
          curves indicate the $1\sigma$ and $2\sigma$ regions for the
          \emph{conservative} configuration.  The analysis assumes that the
          DM mass is known.  In the grey-shaded region the event rate
          in argon is strongly suppressed and therefore the DM mass
          cannot be determined with our method in that region.}
          \label{fig:fnfp}
\end{figure}

For the numerical analysis we have extracted the empirical values
$\hat q_D$ for a large number of random realizations of our benchmark
xenon and argon experiments, assuming a DM mass of 50~GeV, and
calculated the variance of $q_D$ from the samples. Then the precision
with which $\theta$ can be determined from the data can be estimated
with a simple $\chi^2$ analysis, fitting the predicted values $q_1$ and
$q_2$ as a function of $\theta$ to the empirical ones. The result is
shown in Fig.~\ref{fig:fnfp}. We see that for true values far from the
cancellation region around $0.8\pi$, we can exclude that $\theta$ is
close to $0.8\pi$. If the true value is around the cancellation region
and sufficient events are obtained in both detectors that the mass can
be reconstructed, then $\theta$ can be determined very accurately with
our method. This behaviour is obvious from Eq.~\eqref{eq:Aeff-ratio}
and Fig.~\ref{fig:Aeff}.

Note, however, that there is always a degeneracy in the determination
of $\theta$, visible in the plot by the nearly horizontal/vertical
strips. The origin of the degeneracy is related to the sign-ambiguity
in Eq.~\eqref{eq:Aeff-ratio}, since only the ratio of the squares of
the $A_{\rm eff}$ in both experiments can be determined. Therefore,
when solving for $\theta$, there are always two possible sign
combinations, with only one of them corresponding to the true
$\theta$. In the single-isotope approximation the degeneracy is
located at
\begin{equation}
  \tan\theta_{\rm deg} = -
  \frac{2Z_1Z_2 + (Z_1N_2+Z_2N_1)\tan\theta_{\rm true}}
  {Z_1N_2+Z_2N_1 + 2N_1N_2 \tan\theta_{\rm true}} \,,
\end{equation}
with $N_D = A_D-Z_D$, in excellent agreement with the numerical result
in Fig.~\ref{fig:fnfp}. As is clear from the plot, the degeneracy remains
also in the presence of multiple isotopes in xenon.  The only way to
resolve this degeneracy is to consider events in three different
target materials, with sufficiently different proton-to-neutron
ratios. The generalization of our method to three experiments is
straight forward.  For instance, the product of all three form factors
has to be included in the distribution $h(v^2)$ defined in
Eq.~\eqref{eq:h}, and so on. A detailed study of this case is
beyond the scope of this work.

Let us remark that the total cross section $\bar\sigma$ cannot be
extracted halo-independently. DM velocity distribution independent
lower limits on the product $\rho_\chi\bar\sigma$ can be obtained from
averaged rates~\cite{Blennow:2015gta} and, if observed, from annual
modulations~\cite{Herrero-Garcia:2015kga}. The bounds derived there
can be evaluated for the DM mass extracted by applying the method
developed in the present work.

%%%%%%%%%%%%%%%%%%%%%%%%%%%%%%%%%%%%%%%%%%%%%%5
\section{Light mediators\label{sec:light}}
%%%%%%%%%%%%%%%%%%%%%%%%%%%%%%%%%%%%%%%%%%%%%%5

The method explained so far can be directly applied to any differential cross section that is factorizable as the product of velocity and energy-dependent parts. Any extra energy-dependent factor, coming from the differential cross
section or from a DM form factor, can be treated in an analogous way
as the nuclear form factor. As an example, we now assume that the
interaction is mediated by a light mediator, with mass $m_\phi$,
chosen to be in the 10--100~MeV range.  Following the notation of
Ref.~\cite{Schwetz:2011xm}, this situation can be parametrized by an
extra energy-dependent factor in the differential event rate Eq.~\eqref{eq:R}:
\begin{equation}\label{eq:lm}
G_D(E_R) \equiv\frac{(2m_{A_D} E_{{\rm th},D} + m^2_\phi)(2m_{A_D} E_{\rm ref} + m^2_\phi)}{(2m_{A_D} E_R + m^2_\phi)^2}\,,
\end{equation}
where $E_{\rm ref}$ is an arbitrary reference recoil energy, which we
have set to the value corresponding to $v_{\rm m} = 200$~km/s for a
given DM mass according to Eq.~\eqref{eq:vm}.
To take the additional recoil energy dependence into account in our
test, one has to make the replacement
\begin{equation}
F_D^2(E_R) \to F_D^2(E_R)G_D(E_R)  
\end{equation}
in all expressions. Note that a light mediator actually does not
modify the $v_{\rm m}$ distribution itself, but just enters our
analysis in the weight factors. The numerator in Eq.~\eqref{eq:lm}
drops out and we see that the finite mediator mass will only enter
into the CDF in Eq.~\eqref{eq:ecdf} if $2m_{A_D} E_R \simeq m^2_\phi$.
Also, for very light mediators, $2m_{A_D} E_R \gg m^2_\phi$, the
mediator mass is irrelevant and the dependence on the different
targets via $m_{A_D}$ becomes a multiplicative factor which drops
out.\footnote{Note, however, that target-dependent factors are
  important for extracting the relative coupling to neutrons and
  protons, see Eq.~\eqref{eq:Aeff-ratio}. Therefore, information (or
  assumptions) about the mediator mass are important for the analysis
  discussed in section~\ref{sec:fnfp}.\label{foot:lm}} Hence, we expect that our
test, which is only sensitive to non-trivial modifications which are
\emph{different} for the two detectors, will only be sensitive to the
case $2m_{A_D} E_R \simeq m^2_\phi$.

\begin{figure*}[t!]
  \centering
        \includegraphics[height=0.32\linewidth]{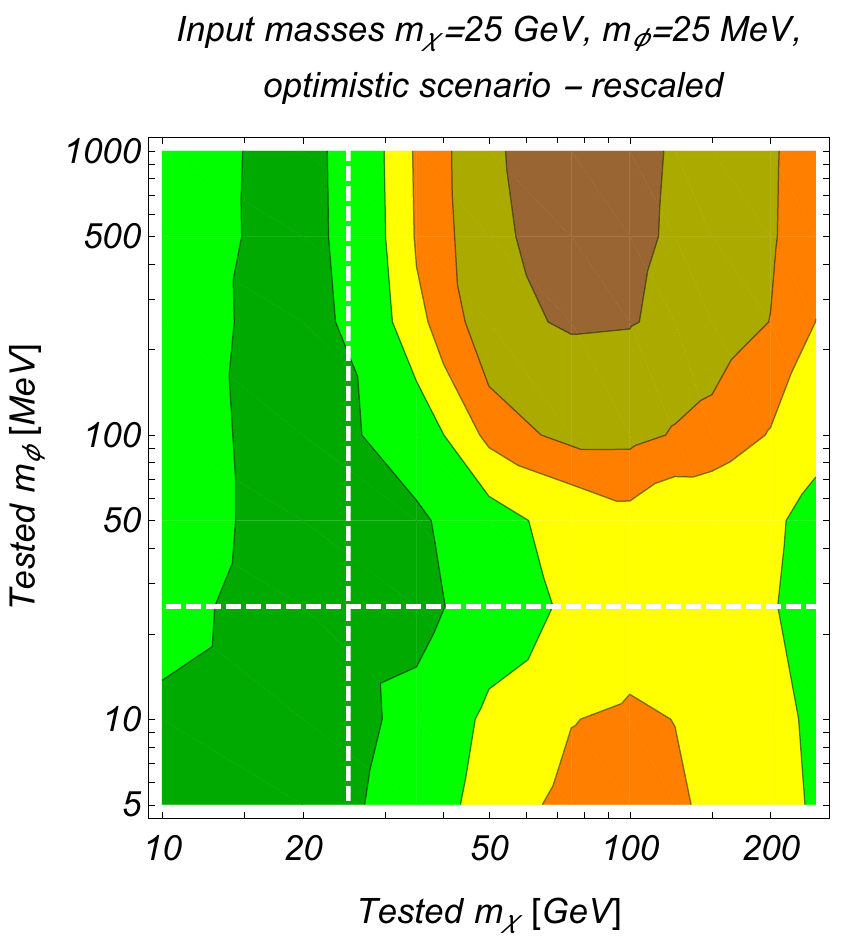}
        \hspace{0.01\linewidth}
	\includegraphics[height=0.32\linewidth]{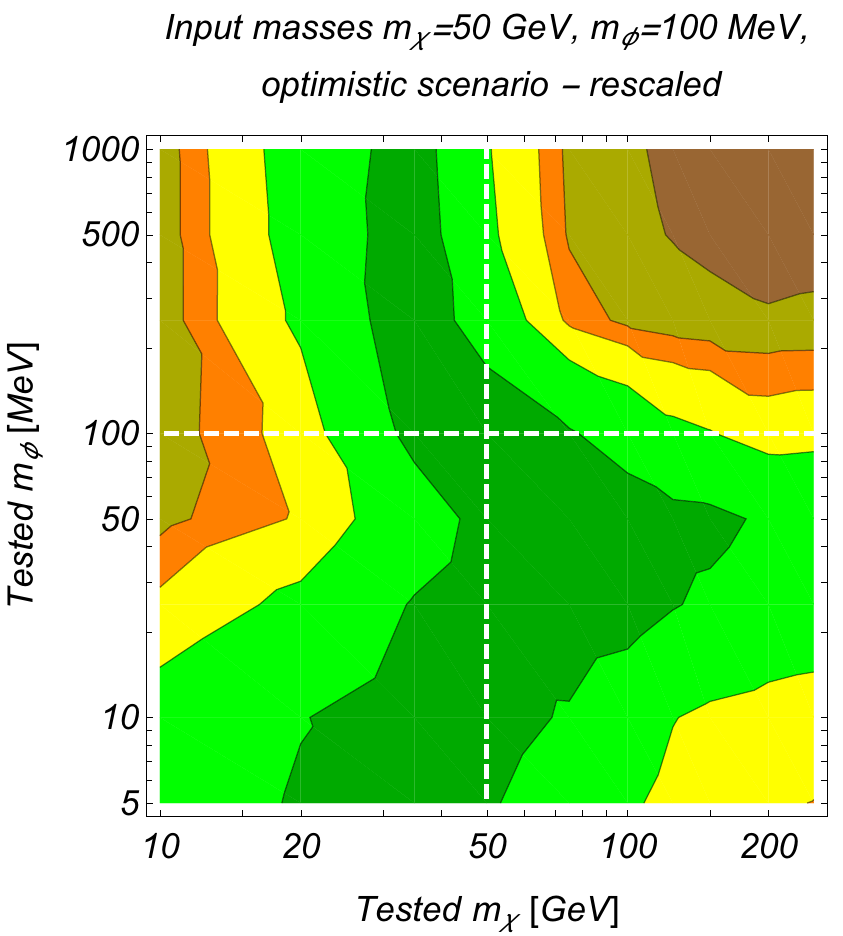}
        \hspace{0.01\linewidth}
	\includegraphics[height=0.32\linewidth]{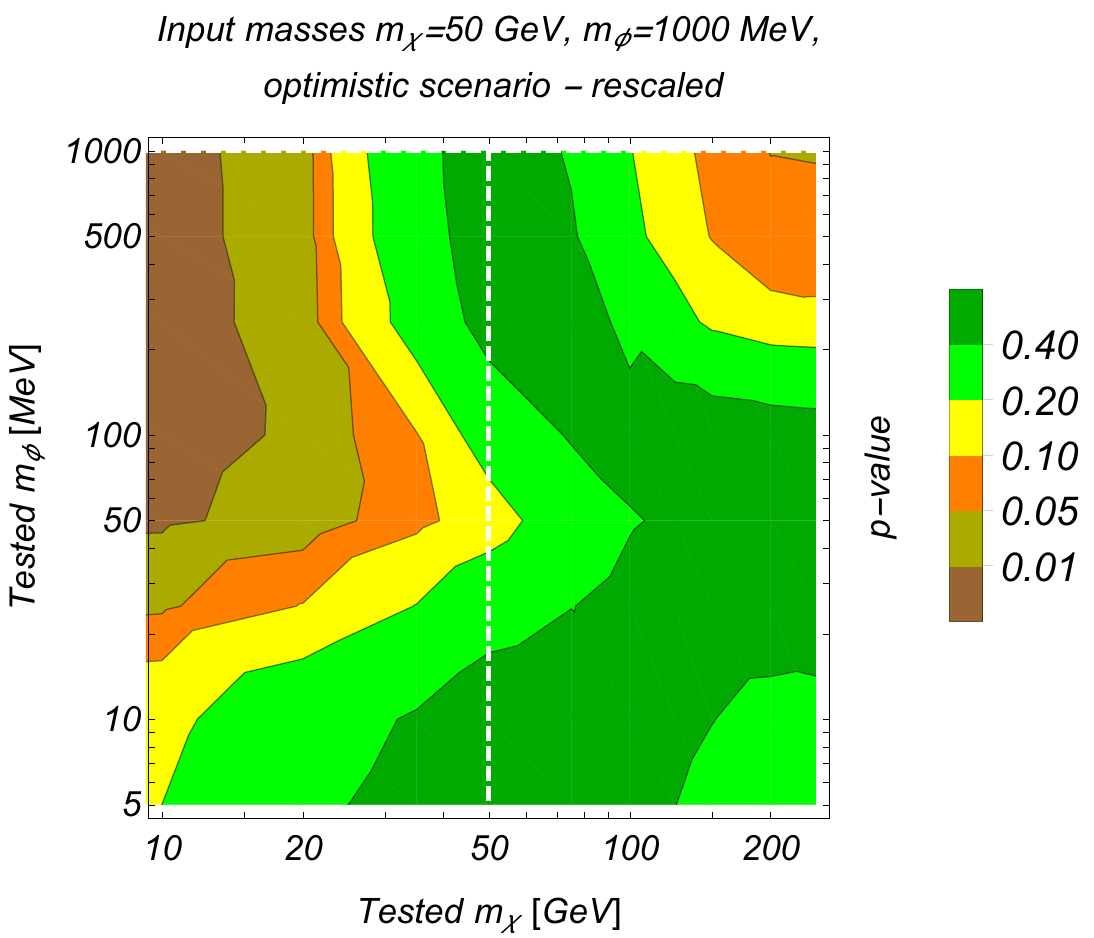}
	\caption{Contours of the median $p$-value for the light
          mediator model in the $m_\phi$ -- $m_\chi$ plane. The input
          values are indicated by the dashed lines and quoted in the
          figure headings. We use the \emph{optimistic} configuration and
          rescale the cross section such that we obtain mean Xe/Ar
          event numbers of 760/187, 770/154 and 849/149 for the left,
          middle and right panels, respectively.} \label{fig:contour_light}
\end{figure*}

In Fig.~\ref{fig:contour_light} we show contour plots of the $p$-value
in the parameter space of tested masses, $m_\phi$ -- $m_\chi$, for
three examples of input masses. We have considered the \emph{optimistic}
scenario and rescaled the total cross section such that event numbers
for our three example points are similar, around 800/160 for Xe/Ar. In
all three cases we see that it is difficult to determine the mediator
mass with our test, for the reasons discussed above, and we observe a
degeneracy between $m_\phi$ and $m_\chi$. For the considered
experimental configurations the condition $2m_{A_D} E_R \simeq
m^2_\phi$ is fulfilled for $m_\phi \simeq 30$~MeV, clearly visible in
the plots.  A value of $m_\phi$ in that region can be compensated by a
larger value of $m_\chi$, while---in agreement with the argument
presented above---the test cannot distinguish between $m_\phi$ much
larger and much smaller than 30~MeV, for similar values of $m_\chi$.
Although event spectra for $m^2_\phi \gg$ or $\ll 2m_{A_D} E_R$ look
quite different, our test is not designed to distinguish between
different spectra itself, but tests differences between the weighted
$v_{\rm m}$ distributions between the two experiments, which indeed
may be identical for the two extreme cases. Clearly additional tests
have to be applied to distinguish those cases, in particular whether
data would be compatible with a physically reasonable halo
model. Despite those limitations of our test, we see from
Fig.~\ref{fig:contour_light} that in all cases certain regions in the
$m_\phi$ -- $m_\chi$ plane can be excluded completely
halo-independently and without any assumption about the neutron-proton
couplings ratio.

%%%%%%%%%%%%%%%%%%%%%%%%%%%%%%%%%%%%%%%%%%%%%%%%%%%%%%%%%%
\section{Conclusions and outlook\label{sec:conclusions}}
%%%%%%%%%%%%%%%%%%%%%%%%%%%%%%%%%%%%%%%%%%%%%%%%%%%%%%%%%%

In the next years significant progress in DM direct detection experiments is to be expected. If finally a DM signal is seen, it is just a matter of time that a signal in a different target detector is also observed. Once this happens, one needs a way to accurately extract the DM parameters by taking into account uncertainties in astrophysical parameters, in particular the  DM velocity distribution and its local energy density. In this work, we have developed a simple halo-independent method to extract the DM mass and the ratio of couplings to neutrons and protons by comparing two DD signals. It is a distribution-free, non-parametric hypothesis test in velocity space, which does not require any binning of the data. Our proposed test is sensitive to the \emph{shape} of the integrated velocity distribution, which has to be identical for the different event samples when converting from nuclear recoil energies to velocity with the correct value of the DM mass.

We have applied the test to mock data from realistic xenon and argon
experiments, such as the DARWIN and DarkSide projects. The test works
best for values of the DM mass between the masses of the two detector
nuclei, which have to be sufficiently different. For heavy DM masses,
only a lower bound can be obtained (as for any method to extract the
DM mass from DM--nucleus scattering data).  For example, a DM mass of
50~GeV can be constrained with our test to the interval [21,\,190]~GeV
at $90\,\%$~CL if 570/100 events are observed in Xe/Ar. If 1200/450
events are available, the interval shrinks to [30,\,90]~GeV. While the
precision is limited, we stress that those results would be completely
independent of any astrophysical assumption, and therefore more robust. Furthermore, we have
presented a method to constrain the ratio of DM couplings to neutrons
and protons, which can be applied to the same data, once the DM mass
has been determined.

A crucial input for the analysis are nuclear form factors. Therefore,
an assumption about the type of interaction is necessary. In our study
we have assumed spin-independent interactions. Generalizations to
other interactions (e.g., spin-dependent) are straight-forward. One
explicit example we have considered is a light mediator particle,
which effectively leads to a modified form factor. In that case we
find that a determination of the mediator mass based on our test is
difficult, however, certain regions in the DM/mediator masses
parameter space can be disfavoured completely halo-independently.

Let us stress that the test uses an absolutely minimal assumption
about the DM distribution, namely that the \emph{shape} of the DM
velocity distribution seen by the two detectors is the same. It is not guaranteed
that the signals are compatible with a physically meaningful DM
distribution. For instance, the requirement that $\eta(v_{\rm m})$ has
to be a decreasing function is not built in the test and should be
checked independently. Once the DM mass has been determined to some
precision, the data can be used to reconstruct the DM distribution, for
instance by methods similar to the ones discussed in the literature,
e.g., \cite{Drees:2007hr,Kavanagh:2013eya,Feldstein:2014ufa}. This
will be an important consistency check, to see whether the data are
consistent with a physically reasonable DM distribution.

Finally, our test or modifications thereof can be applied to a
possible annual modulation signal, to generalized DM--nucleon
interactions with different momentum and/or velocity dependence, to
inelastic scattering, or to multi-component DM. We leave the
exploration of such cases for future work.

\begin{acknowledgments}
	We thank Florian Bernlochner, Paddy Fox and Sam Witte for
        useful discussions. TS would like to thank the Erwin Schr\"odinger
        International Institute, Vienna, for support and hospitality
        during the final stages of this project, and he acknowledges
        support from the European Unions Horizon 2020 research and
        innovation programme under the Marie Sklodowska-Curie grant
        agreement No 674896 (Elusives). JHG acknowledges financial
        support from the H2020-MSCA-RISE project “InvisiblesPlus”, and
        he thanks the Theoretical Physics Department of Fermilab,
        where this project was completed, for hospitality.
\end{acknowledgments}

\bibliography{DM_HI}

\end{document}